\documentstyle [12pt,twoside]{article}

\newcommand{\bq}{\begin{equation}}
\newcommand{\ba}{\begin{eqnarray}}
\newcommand{\eq}{\end{equation}}
\newcommand{\ea}{\end{eqnarray}}
\newcommand{\inti}{\int_{-\infty}^{+\infty}}

\def\a{\alpha}
\def\b{\beta}

\def\d{\delta}
\def\e{\epsilon}

\def\l{\lambda}

\def\n{\nu}

\def\p{\pi}

\def\r{\rho}

\def\L{\Lambda}

\def\bo{{\raise.15ex\hbox{\large$\Box$}}}
\def\bob{{\lower.2ex\hbox{\large$\Box$}}}
\def\pa{\partial}

\def\TH{{\raise.2ex\hbox{$\displaystyle \bigodot$}\mskip-4.7mu \llap H \;}}

\def\VEV#1{\left\langle #1\right\rangle}

\catcode`@=11
\def\underline#1{\relax\ifmmode\@@underline#1\else
        $\@@underline{\hbox{#1}}$\relax\fi}
\catcode`@=12

\begin{document}

\hfill{LA-UR-94-3329}
\centerline{\large{\bf Quantum Diffusion}\normalsize}

\vspace*{1cm}

\centerline{\bf Salman Habib}

\vspace{.5cm}

\centerline{\em T-6, Theoretical Astrophysics}
\centerline{\em and}
\centerline{\em T-8, Elementary Particles and Field Theory}
\centerline{\em Los Alamos National Laboratory}
\centerline{\em Los Alamos, NM 87545}

\vspace*{1cm}

\centerline{\bf Abstract}

{\footnotesize We consider a simple quantum system subjected to a
classical random force. Under certain conditions it is shown that the
noise-averaged Wigner function of the system follows an
integro-differential stochastic Liouville equation. In the simple case
of polynomial noise-couplings this equation reduces to a generalized
Fokker-Planck form. With nonlinear noise injection new ``quantum
diffusion'' terms arise that have no counterpart in the classical
case. Two special examples that are not of a Fokker-Planck form are
discussed: the first with a localized noise source and the other with
a spatially modulated noise source.}

\vspace{1cm}

\noindent{\bf 1. Stochastic Liouville Equations}

Stochastic equations have long been used in physics to model various
phenomena. Brownian motion, spin relaxation, and critical dynamics may
be cited as obvious examples. At a formal level there are two ways to
set up such equations (1) as {\em exact} equations
\cite{RWZ1}\cite{RWZ2} or (2) as part of a phenomenological
description \cite{NVK}. In either case one typically encounters
equations that are nonlocal in time and involve stochastic forcing
terms usually called ``noise.'' Such Langevin equations exist at both
the classical and quantum levels. As expected the situation is more
complicated in the latter case; while in classical problems it is
often possible to approximate the ``noise'' as being Gaussian and
white, and further to replace a nonlocal kernel by one local in time
(the Markov approximation), such simplifications do not easily obtain
in quantum mechanics. Nevertheless, simple approximate approaches are
valuable in that they often capture some essential physics, or even
make some technical point, with less calculational clutter when
compared to a more comprehensive or refined method of attack. The work
outlined here is in this spirit. It owes much to Kubo's study of the
stochastic Liouville equation \cite{KUBO} and a presentation of it
given by Zwanzig \cite{RWZ3}. Different aspects of this work have been
considered in detail elsewhere \cite{SAL}. Nonlinear couplings to an
oscillator environment have been studied in the independent oscillator
model in Ref. \cite{HPZ} where quantum diffusion has also been shown
to exist.

In this paper, all quantum calculations will be done in the Wigner
framework of quantum mechanics. Partly this is because quantum
distribution functions defined on a mock phase space can be easily
compared to their classical counterparts. Furthermore, in the models
that will be discussed, stochastic Liouville equations written in
terms of the Wigner function will be obtained directly from the
stochastic Hamiltonian. This enables us to bypass the somewhat
delicate question of how to derive quantum Fokker-Planck equations
starting from Langevin equations for quantum operators. A nice feature
of the phase space approach is that the quantum derivation of the
stochastic Liouville equation closely parallels the classical
derivation; there is no need to invoke path integrals. Finally, this
approach also enables us to discuss the singular nature of the
$\hbar\rightarrow 0$ limit for both the systematic and the diffusive
terms in the stochastic Liouville equation.

We begin with the Hamiltonian (a generalization of the randomly
forced oscillator considered earlier by Merzbacher \cite{MERZ}):
\begin{equation}
H={p^2\over 2m}+V(x)-F(t)g(x),       \label{one}
\end{equation}
where $p,x$ are the dynamical variables characterizing the motion of
the system. The functions $V(x)$ and $g(x)$ are assumed to be
differentiable. $F(t)$ is an external perturbation that is taken to be
Gaussian, white noise, {\em i.e.}, $\VEV{F(t)}_N=0$, and
\begin{equation}
\VEV{F(t_1)F(t_2)}_N=2B(t_1)\d(t_1-t_2),   \label{two}
\end{equation}
with the usual restrictions on the higher moments. The $\VEV{~~~}_N$
denotes an average over the realizations of $F$. The delta function in
(\ref{two}) is supposed never to be exactly realized, but is treated
just as an idealization of a sharply peaked, symmetric function. This
corresponds to interpreting the noise in the sense of Stratonovich
\cite{RLS}.

One way to write the equations of motion is to use the Liouville
equation for the phase space distribution function. We introduce the
distribution function $f_{Cl}(x,p;t)$, which satisfies the probability
flux conservation equation (Liouville's theorem),
\begin{equation}
{\pa\over \pa t}f_{Cl}(x,p;t)=-{\pa\over\pa x}\left[{\pa H\over\pa
p}f_{Cl}(x,p;t)\right]-{\pa\over\pa p}\left[-{\pa
H\over\pa x}f_{Cl}(x,p;t)\right] ,
\label{five}
\end{equation}
the right hand side of (\ref{five}) defining the Liouville operator
$L_{Cl}$.

Following Kubo's analysis \cite{KUBO} applied to the Hamiltonian
(\ref{one}), we proceed to derive the noise-averaged stochastic
Liouville equation. With $L_0$ the Liouville operator corresponding to
the systematic part of the evolution, we obtain \cite{SAL},
\begin{equation}
{\pa\over\pa t}\VEV{f_{Cl}(t)}_N
=-L_0\VEV{f_{Cl}(t)}_N+\left[B(t)\left({\pa g\over\pa
x}\right)^2\left({\pa^2\over\pa p^2}\right)\right]\VEV{f_{Cl}(t)}_N,
\label{sixteen}
\end{equation}
a Fokker-Planck equation for the noise-averaged distribution function.
Since $f_{Cl}$ is a phase space distribution function, (\ref{sixteen})
is a two-variable Fokker-Planck, or Kramers, equation. In the absence of
noise it reduces to the usual Liouville equation. We observe that
whatever $V(x)$ and $g(x)$ may be, $\VEV{f_{Cl}(t)}_N$ will {\em always}
satisfy a Fokker-Planck equation. This will not be true in the quantum
case, to which we now proceed.

As in the classical case we will work with the stochastic Hamiltonian
(\ref{one}). Because of the noise, this Hamiltonian will evolve pure
states to mixed states. Thus it is appropriate to study not the time
dependent Schr\"{o}dinger equation but rather the quantum Liouville
equation for the density matrix, given here in the coordinate
representation,
\begin{equation}
i\hbar{\pa\over{\pa
t}}\r(x_1,x_2)=\left[H(x_1)-H(x_2)^*\right]\r(x_1,x_2).    \label{twenty}
\end{equation}
We wish to write (\ref{twenty}) in the Wigner formalism of quantum
mechanics \cite{WIG} and then to noise average just as in the
classical case. This derivation is given in the first and third papers
of Ref. \cite{SAL} and here we quote only the final result:
\begin{equation}
{\pa\over\pa t}\VEV{f_W(X,k;t)}_N=-L_{Sys}\VEV{f_W(X,k;t)}_N-\inti
dp~\VEV{f_W(X,k+p;t)}_NK_S(X,p;t),               \label{fifty}
\end{equation}
where
\begin{equation}
K_S(X,p;t)={B(t)\over\p\hbar^3}\inti
dx~\hbox{e}^{2ipx/\hbar}[g(X+x)-g(X-x)]^2   \label{fiftyone}
\end{equation}
and $L_{Sys}$ is the systematic quantum Liouville operator. When $g(X)$
can be profitably Taylor expanded, the above equation can be written
as
\begin{equation}
{\pa\over\pa t}\VEV{f_W(t)}_N
=-L_{Sys}\VEV{f_W(t)}_N+\left[B(t){\bf L}^2\right]\VEV{f_W(t)}_N.
\label{thirtynine}
\end{equation}
where
\begin{eqnarray}
{\bf L}^2&=&\left({\pa g\over \pa X}\right)^2{\pa^2\over\pa
k^2}+2\left({\pa g\over\pa X}\right)
\sum_{\l~{\footnotesize{odd}}}{1\over\l !}\left({\hbar\over
2i}\right)^{\l-1} \left({\pa^{\l}g\over\pa
X^{\l}}\right){\pa^{\l+1}\over\pa k^{\l +1}} \nonumber\\
&&+\sum_{\l,\n~{\footnotesize{odd}}}{1\over\l !\n !}\left({\hbar\over
2i}\right)^{\l +\n -2}\left({\pa^{\l}g\over\pa X^{\l}}\right)
\left({\pa^{\n}g\over\pa X^{\n}}\right){\pa^{\l+\n}\over\pa
k^{\l+\n}}.  \label{fortytwo}
\end{eqnarray}

\noindent{\bf 2. Quantum Diffusion}

The conditions under which (\ref{thirtynine}) will reduce to a
Fokker-Planck form are when both $V(X)$ and $g(X)$ are of the form
$Ax+Bx^2$. In this case the quantum Liouville equation reduces to the
classical one. The difference between the two then lies not in the
dynamical equation, but in the different constraints imposed on the
initial value of the respective distribution functions.

We now study different choices for $g(X)$. If $g(X)=\L X$, with $\L$ a
constant, then ${\bf L}^2=\L^2\pa^2/\pa k^2$, a conventional
diffusion term. This gives rise to the simple model equation often
employed in studies of quantum decoherence \cite{JZ}.

Consider now the case, $g(X)=\L X+{1\over 3}\e X^3$, where
\begin{equation}
{\bf L}^2=(\L +\e X^2)^2 {\pa^2\over\pa k^2}-{1\over 6}(\L +\e
X^2)\e\hbar^2{\pa^4\over\pa k^4} +{1\over
144}\e^2\hbar^4{\pa^6\over\pa k^6}.                \label{fortyfive}
\end{equation}
Notice the appearance of the purely quantum mechanical, higher
even derivative ``diffusion'' terms. The classical limit
$\hbar\rightarrow 0$ is singular not only for the systematic quantum
Liouville operator $L_{Sys}$ \cite{HELL}\cite{SAL2} but also for the
stochastic terms arising from quantum diffusion. It is easy to see
that all the quantum diffusive terms, when acting on ``fast''
(cf. Refs. \cite{HELL}\cite{SAL2}) pieces $\sim \exp(ikX/\hbar)$ of a
Wigner function, are of $O(1/\hbar^2)$. The highest order quantum
diffusion term dominates at large distances and always acts to
increase the linear entropy $1-\int dXdk f^2$ \cite{SAL}. The effect
of the quantum diffusion terms with regard to decoherence is to reduce
the decoherence time at large length scales \cite{SAL}\cite{HPZ}.

\noindent{\bf 3. Two Illustrative Examples}

As we have seen, the stochastic quantum Liouville equation written in
terms of the Wigner distribution function is in general a complicated
integro-differential equation. If the coupling to the noise is through
a polynomial in the system variable, then this equation can truncate
to a finite order partial differential equation. However, there are
cases of physical interest where the coupling to the noise cannot be
reduced to such a form. We will now exhibit two such cases, coupling
the system (1) to a localized noise source, and (2) to a spatially
modulated noise source. The first case is of interest in quantum
tunneling through a stochastic barrier while the second applies to the
noise in a microwave cavity. More details can be found in the third
paper of Ref. \cite{SAL}.

A localized noise source can be modeled by setting $g(X)=\L\exp(-\e
X^2/2)$. In this case,
\ba
{\pa\over\pa
t}\VEV{f_W(X,k;t)}_N&=&-L_{Sys}\VEV{f_W(X,k;t)}_N-{2B\L^2\over
\sqrt{\e\p}\hbar^3} \inti dp \VEV{f_W(X,k;t)}_N   \nonumber\\
&&\times \left[\cos(2pX/\hbar)-\hbox{e}^{-\e
X^2}\right]\hbox{e}^{-p^2/\e\hbar^2}.                   \label{gauss}
\ea

A spatially modulated noise source, $g(X)=\a\sin(\b X/\hbar)$, leads
to
\ba
{\pa\over\pa
t}\VEV{f_W(X,k;t)}_N&=&-L_{Sys}\VEV{f_W(X,k;t)}_N  \nonumber\\
&&-{B\a^2\over\p\hbar^2} \cos^2(\b
X/\hbar)\left[\VEV{f_W(X,k;t)}_N\right.           \nonumber\\
&&\left.-{1\over 2}\VEV{f_W(X,k-\b;t)}_N-{1\over
2}\VEV{f_W(X,k+\b;t)}_N\right]                          \label{mod}
\ea
Eqs. (\ref{gauss}) and (\ref{mod}) are not of a classical form: the
corresponding classical equations result from keeping only the first
term of a derivative expansion of $f_W$ in these equations (the
quantum equations may be viewed as resulting from a resummation of
{\em all} terms in such an expansion).

\noindent{\bf Acknowledgements}\\
I would like to thank Juan Pablo Paz, Bill Unruh, Dan Vollick,
Wojciech Zurek, and especially Robert Zwanzig, for helpful
conversations. This research was supported by the DOE and by AFOSR.


\begin{thebibliography}{99}
\bibitem{RWZ1} R. W. Zwanzig, {\em J. Stat. Phys.} {\bf 9}, 215 (1973).
\bibitem{RWZ2} R. W. Zwanzig, in {\em Systems Far from Equilibrium},
edited by L. Garrido (Springer, New York, 1980).
\bibitem{NVK} N. G. van Kampen, {\em Stochastic Processes in
Physics and Chemistry} (North-Holland, New York, 1981).
\bibitem{KUBO} R. Kubo, {\em J. Math. Phys.} {\bf 4}, 174 (1963).
\bibitem{RWZ3} R. W. Zwanzig, (unpublished).
\bibitem{SAL} S. Habib, UBC Report (1990) (unpublished); {\em
Phys. Rev. D} {\bf 46}, 2408 (1992); Los Alamos preprint (1994).
\bibitem{HPZ} B. L. Hu, J. P. Paz, and Y. Zhang, {\em Phys. Rev. D}
{\bf 47}, 1576 (1993).
\bibitem{MERZ} E. Merzbacher, {\em Physica} {\bf 96A}, 263 (1979).
\bibitem{RLS} R. L. Stratonovich, {\em Conditional Markov Processes
and Their Application to the Theory of Optimal Control} (Elsevier, New
York, 1968).
\bibitem{WIG} E. P. Wigner, {\em Phys. Rev.} {\bf 40}, 749 (1932).
\bibitem{JZ} E. Joos and H. D. Zeh, {\em Z. Phys. B} {\bf 59}, 223
(1985).
\bibitem{HELL} E. J. Heller, {\em J. Chem. Phys.} {\bf 65}, 1289
(1976).
\bibitem{SAL2} S. Habib, {\em Phys. Rev. D} {\bf 42}, 2566 (1990).
\end{thebibliography}
\end{document}